\documentclass[manuscript]{emulateapj}
\usepackage{lscape,graphicx,rotating}

\slugcomment{Draft Version - March 31, 2009}

\newcommand{\diaz}{N$_2$H$^+$}
\newcommand{\rto}{$\rightarrow$}

\shorttitle{Submillimeter Observations of Ophiuchus A-N6}
\shortauthors{Pon et al.}

\begin{document}

\title{Submillimeter Observations of the Quiescent Core - Ophiuchus A-N6}

\author{A. Pon\altaffilmark{1,2,3}, R. Plume\altaffilmark{2,4}, R. K. Friesen\altaffilmark{1,3}, J. Di Francesco\altaffilmark{3,1,2}, B. Matthews\altaffilmark{3,1}, \& E. A. Bergin\altaffilmark{5}}
 
\altaffiltext{1}{Department of Physics and Astronomy, University of Victoria, PO Box 3055 STN CSC, Victoria BC V8W 3P6}
\altaffiltext{2}{Centre for Radio Astronomy, University of Calgary, 2500 University Dr. NW, Calgary, AB, T2N 1N4, CA}
\altaffiltext{3}{NRC-Herzberg Institute of Astrophysics, 5071 W. Saanich Road, Victoria, BC, V9E 2E7, CA}
\altaffiltext{4}{Max Planck Institute for Astronomy, K\"{o}nigstuhl 17, 69117 Heidelberg, Germany}
\altaffiltext{5}{Department of Astronomy, University of Michigan, 825 Dennison Building, Ann Arbor, MI 48109}

\begin{abstract}

We have observed the Oph A-N6 prestellar core in the following transitions: N$_2$D$^+$ J=3\rto2, DCO$^+$ J=3\rto2 and J=5\rto4, HCO$^+$ J=3\rto2, CS J=5\rto4 and J=7\rto6, and H$^{13}$CO$^+$ J=3\rto2 and J=4\rto3, using the James Clerk Maxwell Telescope.  We also observed the NH$_3$ (1,1) and (2,2) inversion transitions towards the Oph A-N6 peak with the Green Bank Telescope. We have found that the N6 core is composed of shells of different chemical composition due to the freezing out of chemical species at different densities. The undepleted species N$_2$D$^+$ appears to trace the high-density interior of the core, DCO$^+$ and H$^{13}$CO$^+$ trace an intermediate region, and CS traces the outermost edges of the core. A distinct blue-red spectral asymmetry, indicative of infall motion, is clearly detected in the HCO$^+$ spectra, suggesting that N6 is undergoing gravitational collapse. This collapse was possibly initiated by a decrease in turbulent support suggested by the fact that the non-thermal contribution to the line widths is smaller for the molecular species closer to the center of the core. We also present a temperature profile for the core. These observations support the claim that the Oph A-N6 core is an extremely young prestellar core, which may have been recently cut off from MHD support and begun to collapse.  	
	
 \end{abstract}
 

\section{INTRODUCTION}
\label{intro}

	Giant molecular clouds (GMCs) are well known to be the birthplaces of stars. Studies of various GMCs have not only revealed numerous protostellar objects of varying age, they have also detected localized maxima in dust column density that exhibit no signs of star formation and have been designated as starless cores (e.g., Motte et al. 1998; Andr\'{e} et al. 1993; J\o rgensen et al. 2007; Di Francesco et al. 2007). Some of these starless cores have been shown to be gravitationally bound and have been labeled as prestellar cores, as they are believed to be the progenitors of protostars (e.g., Di Francesco et al 2007; Andr\'{e} et al. 2007 and references therein). Further examination of prestellar cores can shed light on the earliest stages of star formation and, in particular, the processes that lead to the gravitational collapse of a core.

	The Ophiuchus complex is particularly suitable for studying star formation as it is relatively close at only 120 pc (Loinard et al. 2008) and is an active star forming region (Loren et al. 1990). Previous studies have found multiple clumps inside the Ophiuchus complex (Loren et al. 1990, Johnstone et al. 2004). Of particular interest is the Oph A clump, the brightest of the clumps at mm wavelengths (Motte et al. 1998), in which numerous starless cores and the prototypical Class 0 object, VLA 1623 (Andr\'{e} et al. 1993), are located. Class 0 objects are deeply embedded protostars that have yet to accrete the majority of their final masses (Barsony 1994). The Oph A clump has previously been surveyed in millimeter wavelengths (e.g. Andr\'e et al. 1993) and in numerous spectral lines including C$^{18}$O (J=1\rto0) (Wilking \& Lada 1983), $^{13}$CO (J=1\rto0) (Loren 1989), CS (J = 2\rto1) (Liseau et al. 1995), and DCO$^+$ (J=2\rto1, 3\rto2) (Loren et al. 1990).

	Recently, the N$_2$H$^+$ J=1\rto0 emission from the Oph A clump was mapped with the BIMA array and the IRAM 30m telescope (Di Francesco et al. 2004, hereafter DAM04). DAM04 discovered that the maxima of N$_2$H$^+$ emission are not coincident with the maxima of 1.3 mm continuum emission found by Motte et al. (1998). This result was inconsistent with results from other low-mass starless cores surveyed by Tafalla et al. (2002). Within Oph A, VLA 1623 was found to have very little N$_2$H$^+$ emission, SM2, a previously identified starless core (Andr\'e et al. 1993), was found to have weak N$_2$H$^+$ emission and a new core not identified in Motte et al.'s (1998) millimeter continuum survey as a continuum maximum, designated Oph A-N6, was found as an emission peak in N$_2$H$^+$. DAM04 suggested that the difference in N$_2$H$^+$ emission from these three cores was due to the cores being in different stages of development with N6 being the youngest and VLA 1623 being the oldest. 
	
\begin{figure}[ht] 
   \centering
  \includegraphics[angle=-90, width=6in]{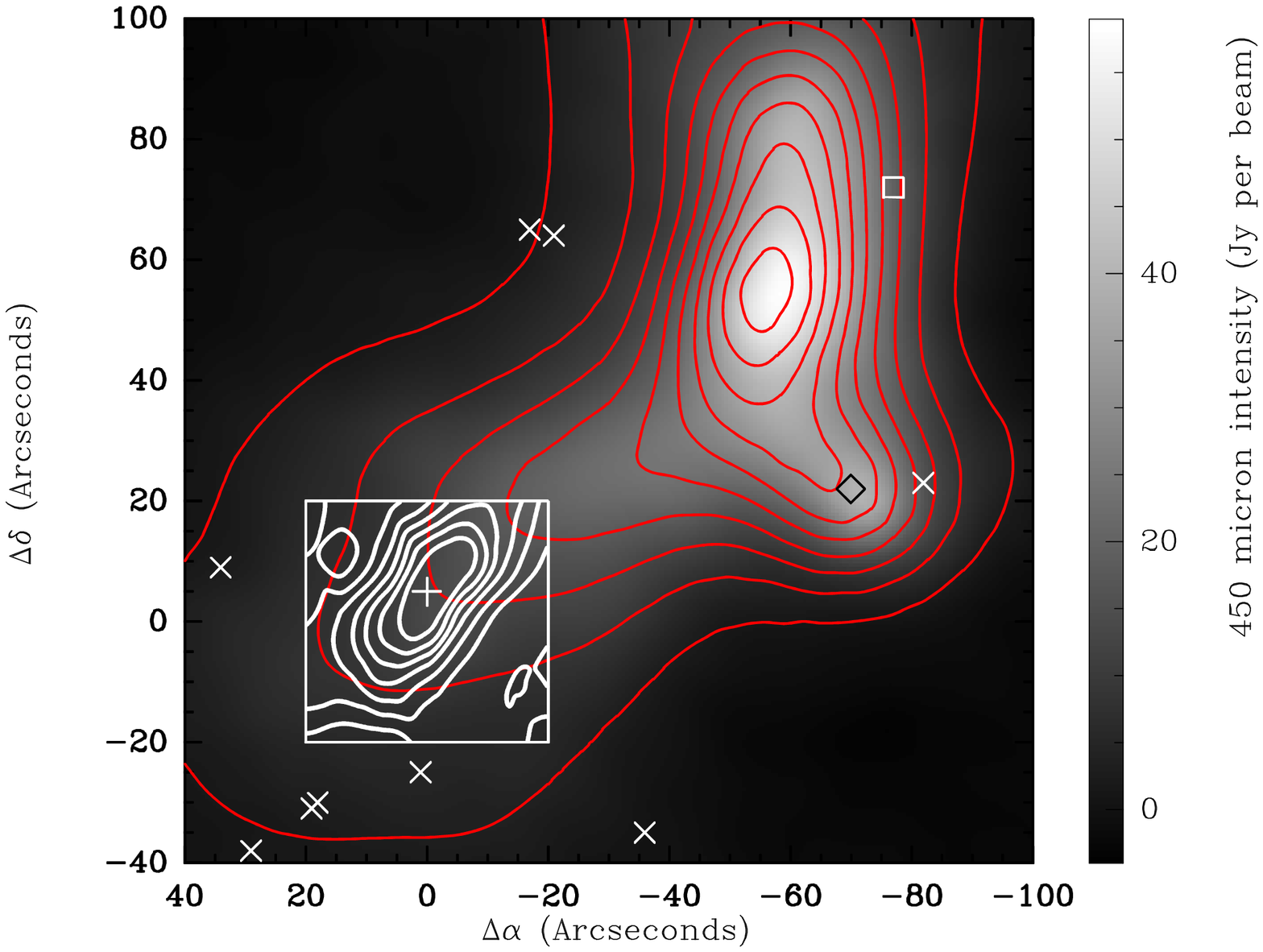} 
   \caption{The grey scale shows the 450 $\mu$m intensity while the red contours show the 850 $\mu$m intensity from the Oph A clump.  Contours are separated by 10\% of the maximum 850 $\mu$m intensity. Both of these data sets came from the SCUBA Legacy Catalogues (Di Francesco et al. 2008). Note how this dust continuum emission exhibits a northwest to southeast gradient across our standard grid, which is outlined by the large white box. The integrated intensity of the N$_2$H$^+$ 1\rto0 transition around the N6 core, as observed by DAM04, is shown as the white contours which are spaced by twice the uncertainty of 1.4 K km s$^{-1}$ estimated by DAM04. Enoch et al. (2008) suggest that the objects in the c2d catalog that are most likely to be protostellar objects are those denoted as either red or YSOc (young stellar object candidate) objects. The locations of all red (Xs), red1 (diamonds) and YSOc (small squares) objects in the c2d catalog near Oph A have been marked on the figure. Note that no Spitzer source likely to be a protostellar object is found within our standard grid. The cross denotes the location of the submillimeter continuum peak found interferometrically by Bourke et al. (2009, in preparation).}
   \label{fig:dust}
\end{figure}

	No intensity peak coincident with the location of the N6 core was found in either the 850 $\mu$m or the 450 $\mu$m surveys of Oph A conducted with the Submillimeter Common-User Bolometric Array (SCUBA) on the James Clerk Maxwell Telescope (Wilson et al. 1999; Johnstone et al. 2000). These SCUBA data were later re-reduced and released as part of the SCUBA Legacy Catalogues (Di Francesco et al. 2008). An interferometric submillimeter survey by Bourke et al. (2009, in preparation), however, detected a localized intensity maximum at the location of the N6 core. No mid-infrared point source, down to 0.001 L$_\odot$, consistent with the presence of a protostar was found within the boundaries of the N6 core by the Spitzer Cores to Disks (c2d) survey (Evans et al. 2003, Padgett et al. 2008), suggesting that the N6 core most likely does not contain a protostar. Figure \ref{fig:dust} shows the 850 $\mu$m and 450 $\mu$m emission from the Oph A clump (Di Francesco et al. 2008), the N$_2$H$^+$ emission from the N6 core (DAM04), the location of the submillimeter intensity maximum found by Bourke et al. (2009, in preparation) and the locations of Spitzer sources consistent with the presence of protostars in the Oph A clump as identified by the c2d survey (Evans et al. 2003). 

	DAM04 found that the \diaz\ line widths in N6 were only slightly larger than expected from thermal motions alone and that the N6 core is, to first order, in gravitational virial equilibrium. Thus, all evidence suggests that the N6 core is a prestellar core and that it is no longer strongly interacting with the turbulent flows of its surroundings. It has been suggested that turbulent motions could be reduced in dense cores if the cores become smaller than the cut-off wavelength for magnetohydrodynamic (MHD) waves in the region (e.g., McKee et al. 1993; Nakano 1998; Myers 1998). Since turbulent gas flows and magnetic support are likely important parts of a core's support against gravitational collapse (Caselli et al. 2002; Myers 1999; McKee 1989), the N6 core may be on the verge of gravitational collapse. This makes the N6 core a valuable object for further study into the removal of support against gravitational collapse. 
	
	In Section \ref{obs}, we describe the details of our molecular line observations of the N6 core and the results of these observations are discussed in Section \ref{results}. We also present derived dust temperatures and column densities of N$_2$D$^+$ in Section \ref{results}. We then propose a model for the chemical and physical state of the N6 core based upon these observations in Section \ref{discuss}. We present a summary of our conclusions in Section \ref{conclusions}.

\section{OBSERVATIONS}
\label{obs}

\subsection{Submillimeter Data}
\label{jcmt}

	The following observations were obtained at the James Clerk Maxwell Telescope (JCMT) on Mauna Kea using the position-switching mode of the telescope. Data reduction was performed using a combination of the JCMT's STARLINK software and the CLASS data reduction package.

	Between 2005 February and 2005 July, observations were made of the N$_2$D$^+$ J=3\rto2, DCO$^+$ J=3\rto2, DCO$^+$ J=5\rto4, HCO$^+$ J=3\rto2, and CS J=7\rto6 transitions. Between 2007 March and 2007 June, observations of the CS J=5\rto4, H$^{13}$CO$^+$ J=3\rto2, and H$^{13}$CO$^+$ J=4\rto3 transitions were made. Table \ref{tab:obs} gives the rest frequencies of each of the above transitions along with the front and back end instruments used, the spectral resolutions obtained, the full width at half maximum (FWHM) of the beams, the beam efficiencies, and the baseline RMS achieved for these observations. The critical densities of these transitions are also given. 
	
	All observations, except those of the H$^{13}$CO$^+$ J=4\rto3 transition, were made on a 5 x 5 grid with 10\arcsec\ spacing centered on the position of maximum N$_2$H$^+$ J=1\rto0 integrated intensity found by DAM04, i.e. [RA, decl.] = [$16^h26^m31.6^s$, -$24\,^{\circ}24\arcmin52\arcsec$] (J2000), which we define as our standard grid. Unless otherwise stated, all positional offsets in this paper are relative to the central point of this standard grid and are given in arcseconds. We will consider this position of maximum N$_2$H$^+$ integrated intensity to be the central location of the N6 core. The H$^{13}$CO$^+$ J=4\rto3 emission was observed on a 16 x 16 grid with 7.5\arcsec\ spacing. 
	
	Main beam efficiencies  ($\eta_{mb}$) were obtained from observations of Mars and Jupiter, using the method detailed by Mangum (1993), which accounts for the coupling of a planetary disk with a Gaussian beam.

\subsection{Ammonia Data}
\label{gbt}

	Observations of NH$_3$ (1,1) and (2,2) inversion line emission, at a single pointing centered on the N$_2$H$^+$ maximum position, were obtained using the 100 m Robert C. Byrd Green Bank Telescope (GBT). A total of ten minutes was spent on source over 2006 April 9-10 while the source was at $\sim 22^\circ$ elevation and $T_{sys} = 45 - 55$ K. The observations were done in frequency-switched mode using the GBT K-band (upper) receiver as the front end and the GBT spectrometer as the backend. This setup allowed for the observation of both lines simultaneously in two 50\,MHz-wide IFs, each with 8192 spectral channels, giving a frequency resolution of 6.104\,kHz, or 0.077\,km\,s$^{-1}$ at 23.694\,GHz. The average telescope aperture efficiency $\eta_A$ and main beam efficiency $\eta_{mb}$ were $0.59 \pm 0.05$ and $0.78 \pm 0.06$ respectively, and were determined through observations of the recommended calibration source 3C286 at the start of each shift. The absolute flux accuracy is thus $\sim 8$\,\% and the FWHM beam size for these observations was approximately 32\arcsec.

All data in this paper are presented in units of T$_A$$^*$ (Kutner \& Ulich 1981).  However, to compare our observations to model predictions and to calculate physical conditions, we included the main beam coupling efficiency ($\eta_{mb}$) such that T$_{mb}$ = $T_A^*/\eta_{mb}$.  

\section{RESULTS}
\label{results}

\subsection{Integrated Intensity Maps}
\label{morph}

\begin{figure}[ht] 
   \centering
   \includegraphics[angle=-90, width=6in]{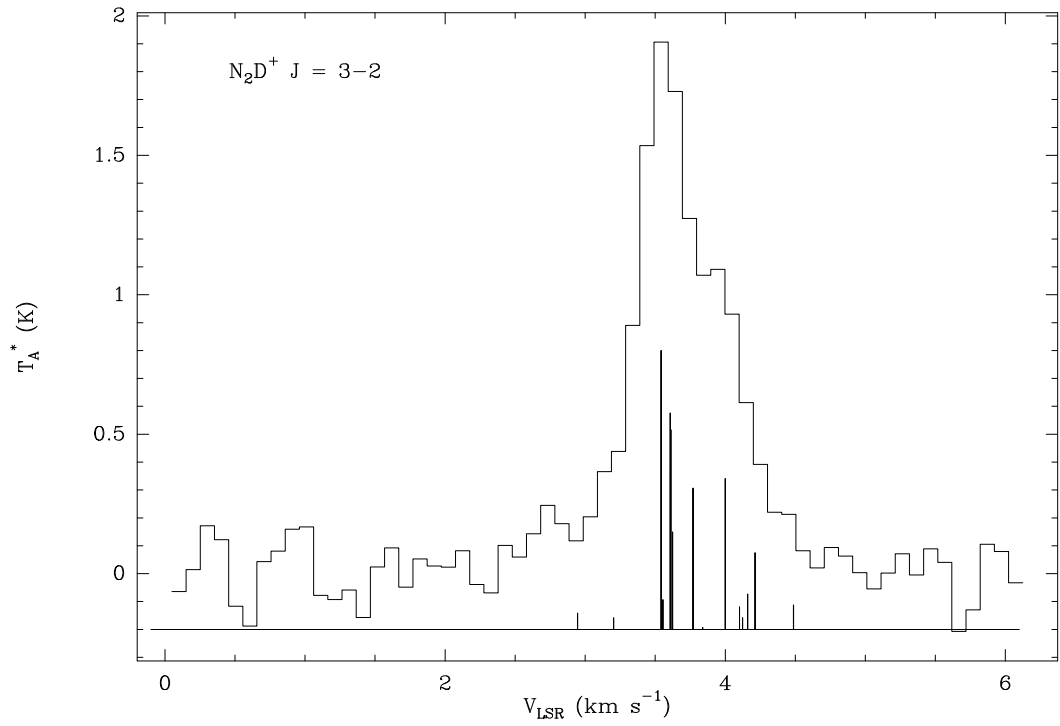} 
   \caption{The observed N$_2$D$^+$ 3\rto2 spectrum at the position of maximum N$_2$D$^+$ integrated intensity, the [-10,10] position, is shown as the histogram. The lower spectrum shows the expected positions and relative strengths of the 16 hyperfine components of this transition.}
   \label{fig:n2dpspec}
\end{figure}

	Figure \ref{fig:allmol} shows integrated intensity maps for the N$_2$H$^+$ 1\rto0 transition observed by DAM04 and for the N$_2$D$^+$ 3\rto2, CS 5\rto4, CS 7\rto6, DCO$^+$ 3\rto2, DCO$^+$ 5\rto4, H$^{13}$CO$^+$ 3\rto2, and H$^{13}$CO$^+$ 4\rto3 transitions that we observed. For the CS, DCO$^+$ and H$^{13}$CO$^+$ transitions, the integrated intensities were found by fitting Gaussian curves to the observed spectra. The observed N$_2$D$^+$ spectra are composites of at least 16 strong hyperfine components (Gerin et al. 2001) and these hyperfine components are badly blended (as shown in Figure \ref{fig:n2dpspec}), so we have calculated the integrated intensity of the N$_2$D$^+$ 3\rto2 spectra by summing all emission from $V_{LSR}$ = 2.1 km s$^{-1}$ to $V_{LSR}$ = 5 km s$^{-1}$ after baseline subtraction. We rejected any spectra that did not have a peak intensity at least 3x the 1$\sigma$ (RMS) noise of each transition (see Table \ref{tab:lineprop}). We estimated the average uncertainty in the integrated intensity of a transition by multiplying the average baseline RMS by the average width of the innermost 9 spectra of that transition. 
	
	The N$_2$D$^+$ distribution, while showing a slight elongation in the northwest-southeast direction, appears to be less elongated than the N$_2$H$^+$ distribution. While the peak integrated intensity for the N$_2$D$^+$ occurs at the [-10,10] position, there is still a small plateau at the central [0,0] position where the N$_2$H$^+$ peaks in intensity.  Figure \ref{fig:n2dpspec} shows the N$_2$D$^+$ 3\rto2 spectrum at the [-10,10] position. The N$_2$D$^+$ integrated intensity peak occurs only 3\arcsec, or 15\% of the beam size, from the location with the narrowest N$_2$H$^+$ line widths (0.19 km s$^{-1}$) as reported by DAM04.
		
	Little CS emission is seen at the positions of peak N$_2$H$^+$ and N$_2$D$^+$ integrated intensity.  Instead, the CS integrated intensity maps show emission in a ring-like structure around these positions. 
		
	The DCO$^+$ integrated intensity maps also show little sign of the N6 core that is seen so clearly in the \diaz\ and N$_2$D$^+$ maps. Instead, the DCO$^+$ maps show decreasing brightness gradients from northwest to southeast and closely resemble the dust continuum profiles shown in Figure \ref{fig:dust}. The only indication of the N6 core in the DCO$^+$ data is the presence of slight plateaus in the brightness gradient close to the [0,0] and [-10,10] positions. 

	Like the DCO$^+$, the H$^{13}$CO$^+$ integrated intensity maps show little sign of the N6 core and are, instead, dominated by single brightness gradients. The H$^{13}$CO$^+$ brightness gradients, however, lie along west-to-east lines rather than the northwest-to-southeast direction seen in the DCO$^+$ data. The plateaus seen in the DCO$^+$ data are also not present in the H$^{13}$CO$^+$ data. 
		
\subsection{Line Widths and Centroid Velocities}
\label{vlsr}	

	Table \ref{tab:lineprop} lists the average line widths and centroid velocities for the innermost nine pointings of the N$_2$D$^+$ 3\rto2, CS 5\rto4, CS 7\rto6, DCO$^+$ 3\rto2, DCO$^+$ 5\rto4, H$^{13}$CO$^+$ 3\rto2, and H$^{13}$CO$^+$ 4\rto3 transitions. We have regridded the H$^{13}$CO$^+$ 4\rto3 data onto our standard grid for this analysis and we have used CLASS' hyperfine splitting routine to determine the line widths and centroid velocities for the N$_2$D$^+$ data.  
	
	The N$_2$D$^+$ line widths tend to be the smallest while the H$^{13}$CO$^+$ and CS line widths tend to be the largest. The average N$_2$D$^+$ line width of 0.30 km s$^{-1}$ is similar to the average N$_2$H$^+$ 1\rto0 line width of 0.25 km s$^{-1}$ reported by DAM04 and 0.37 km s$^{-1}$ reported by Andr\'{e} et al. (2007), who used a 26\arcsec\ FWHM beam. While there appear to be significant differences in average centroid velocity between the different transitions, we found that there were discrepancies between the rest frequencies given in the JCMT database, which we have used, the Leiden Atomic and Molecular database (LAMDA) and the NASA Jet Propulsion Laboratory (JPL) database for some of the above transitions. These discrepancies were on the order of 0.1 to 0.2 km s$^{-1}$. Consequently, we cannot say whether these centroid velocity differences are significant. The individual transitions do not show significant point-to-point variations in centroid velocity.
			
\subsection{Line Asymmetry in HCO$^+$}
\label{hco}
	
\begin{figure}[ht] 
   \centering
  \includegraphics[angle=-90, width=7in]{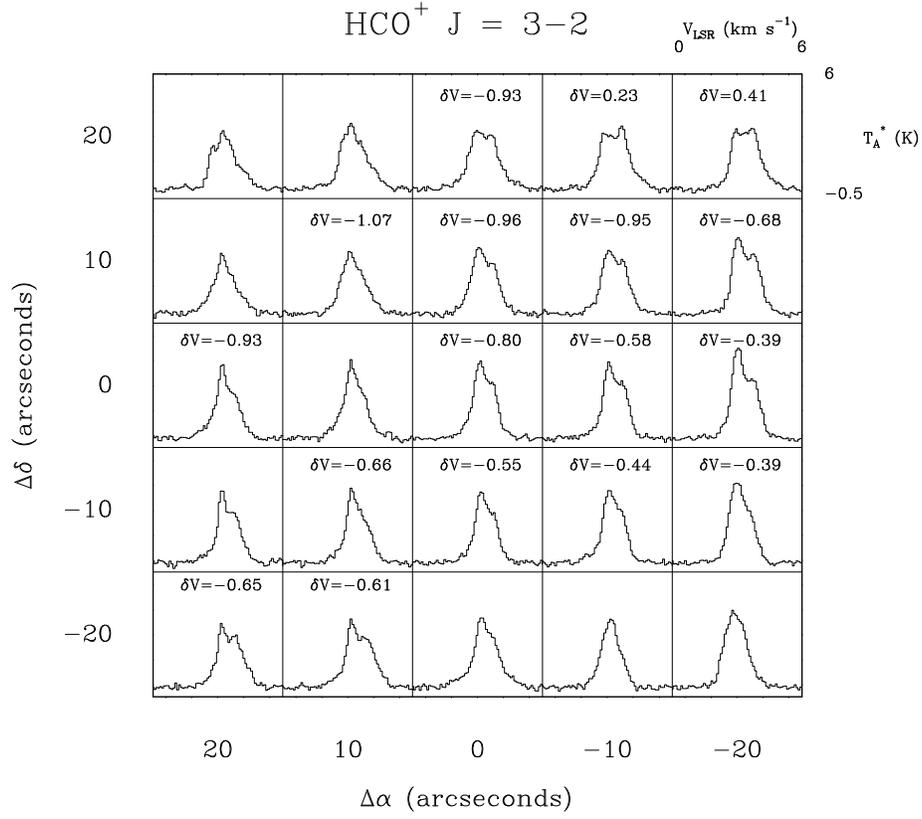} 
   \caption{Grid of HCO$^+$ 3\rto2 spectra around the N6 core. The abscissa of each spectrum is $V_{LSR}$ and ranges from 0 km s$^{-1}$ to 6 km s$^{-1}$. The ordinate of each spectrum is $T_A^*$ and ranges from -0.5 K to 6 K. Note the double peak structure of many of the spectra. The asymmetry parameter, $\delta V$, for all spectra to which we were able to fit Gaussians is also shown.}
      \label{fig:hcop}
\end{figure}

\begin{figure}[ht] 
   \centering
   \includegraphics[angle=-90, width=6in]{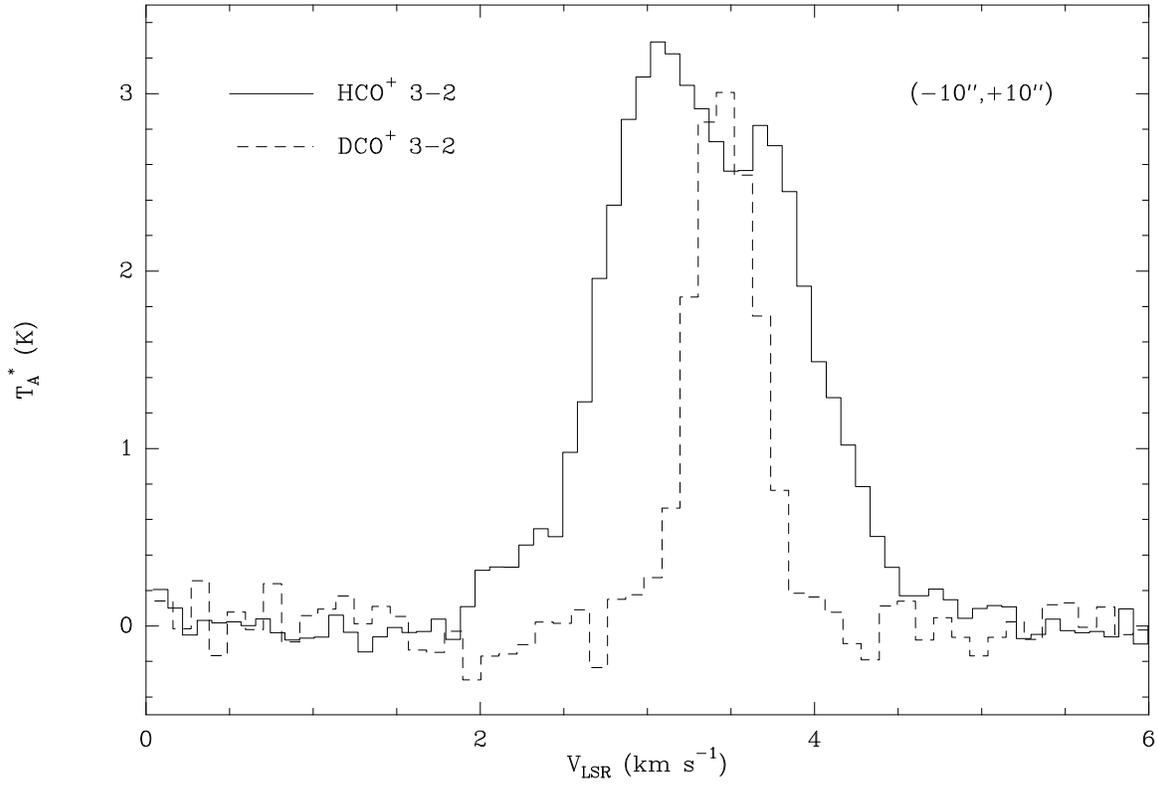} 
   \caption{HCO$^+$ 3\rto2 (solid line) and DCO$^+$ 3\rto2 (dashed line) spectra at the location of peak N$_2$D$^+$ intensity ([-10,10] location). Note the alignment of the DCO$^+$ line with the absorption feature in the HCO$^+$ spectrum and the blue asymmetry in the HCO$^+$ spectrum indicative of infall (e.g., Gregersen et al. 1997; Myers et al. 2000).}
   \label{fig:infall}
\end{figure}

	The spectra from CS, DCO$^+$, and H$^{13}$CO$^+$ show symmetric profiles with no self-absorption. Figure \ref{fig:hcop}, however, shows that the HCO$^+$ 3\rto2 spectra have asymmetric profiles. The HCO$^+$ spectra located near the central position of the N6 core show blue peaks around $V_{LSR}$ = 3.1 km s$^{-1}$, absorption features around $V_{LSR}$ = 3.4 km s$^{-1}$ and smaller red peaks around $V_{LSR}$ = 3.8 km s$^{-1}$. Figure \ref{fig:infall} shows an example of how the optically thin DCO$^+$ 3\rto2 line appears at the same velocity as the absorption feature for the HCO$^+$ 3\rto2 spectra at the same location. This correspondence is seen in all HCO$^+$ 3\rto2 spectra that have absorption features. A few locations along the $\delta=+20$ row, see Figure \ref{fig:hcop}, appear to show reverse asymmetry.	
	
	 We have calculated the asymmetry parameter suggested by Mardones et al. (1997) for the 17 spectra for which we were able to simultaneously fit Gaussian profiles to both the red and blue peaks. The asymmetry parameter, $\delta V$, is calculated as:
\begin{equation}
\delta V = \frac{V_{thick}-V_{thin}}{\Delta V_{thin}}
\end{equation}
where $V_{thick}$ is the centroid velocity of the brightest peak in the optically thick spectrum (HCO$^+$ 3\rto2), $V_{thin}$ is the centroid velocity of the optically thin spectrum (DCO$^+$ 3\rto2) and $\Delta V_{thin}$ is the line width of the optically thin spectrum. The asymmetry parameters for these 17 spectra are listed on Figure \ref{fig:hcop}.

	For the 15 spectra that have a larger blue peak, every asymmetry parameter was calculated to be more negative than $-0.25$, which places these spectra into the ``significant blue shift'' category of Mardones et al. (1997), and the average asymmetry parameter for these 15 locations was calculated to be $-0.71$. For the two spectra with larger red peaks, the [-10,20] position had an asymmetry parameter that showed ``no significant shift'' while the asymmetry parameter at [-20,20] indicated a ``significant red shift''. While both the [0,0] and [-10,10] positions, the positions of peak N$_2$H$^+$ and N$_2$D$^+$ column densities respectively, show significant blue shifts, the asymmetry parameters for these two locations are not significantly different from the other 13 locations that have significant blue shifts. 
		
	Mardones et al. (1997) also defined a blue excess value, E, as:
\begin{equation}
E=\frac{N_{-} - N_{+}}{N} 
\end{equation}
where $N_{-}$ is the number of spectra with a significant blue shift, $N_{+}$ is the number of spectra with a significant red shift and N is the total number of spectra. We find a blue excess value of 0.52 for the N6 core, which is greater than the significance threshold value of 0.25 of Mardones et al. (1997).   Interpretation of these results will be presented in Section \ref{infall}.
				
\subsection{Temperature Structure}
\label{temp}

\begin{figure}[ht] 
   \centering
   \includegraphics[angle=0, width=6in]{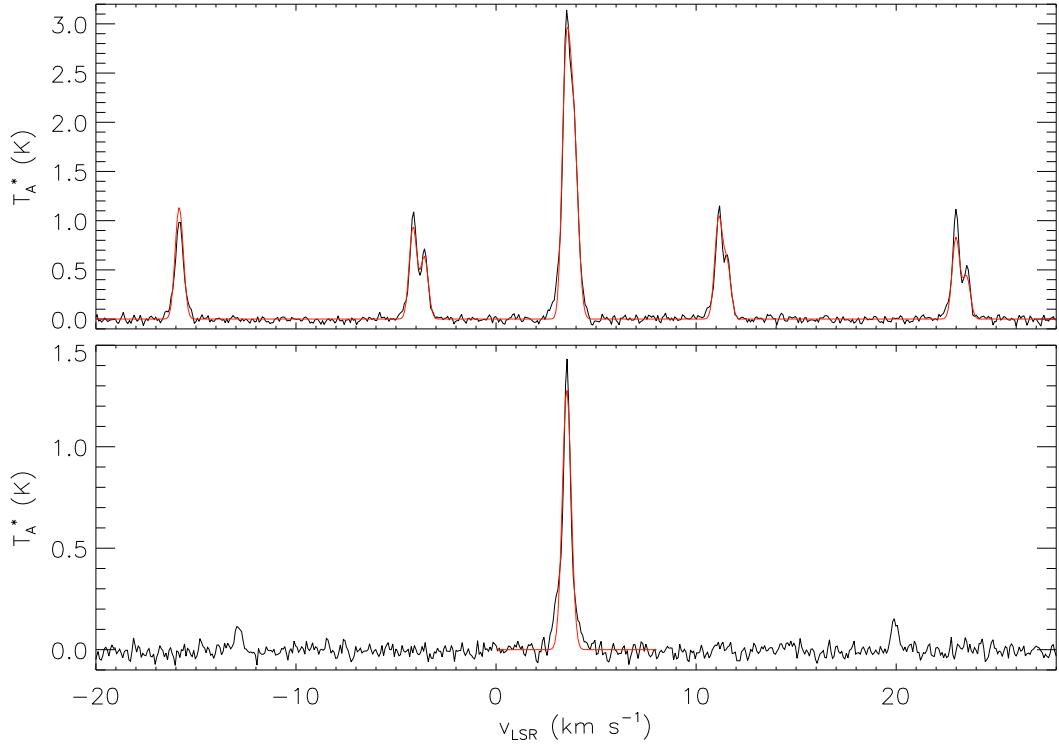} 
   \caption{The top spectrum shows the NH$_3$ (1,1) transition from the [0,0] position in black and our best fit to the spectrum in red. The lower spectrum shows the NH$_3$ (2,2) transition from the [0,0] position in black and our best Gaussian fit in red.}
   \label{fig:nh3}
\end{figure}

	Figure \ref{fig:nh3} shows the NH$_3$ spectrum from the [0,0] position. Assuming Gaussian profiles, the 18 hyperfine components of the NH$_3$ (1,1) line shown in Figure \ref{fig:nh3} were fit simultaneously using a chi-square reduction routine custom written in {\sc IDL} [see Friesen et al. (2009)]. The returned fit, overlaid on Figure \ref{fig:nh3}, provide estimates of the line centroid velocity, $V_{LSR}$, the line FWHM, $\Delta V$, the opacity of the line summed over the 18 components, $\tau_{tot}$, and $[J(T_{ex}) - J(T_{bg})]$. As shown in Figure \ref{fig:nh3}, the satellite components of the NH$_3$ (2,2) line are visible above the rms noise of our data. They are not detected, however, with sufficient signal-to-noise to fit the full hyperfine structure of the transition, and the main component seen in these data was consequently fit with a single Gaussian. Assuming LTE, we then determined the rotational temperature $T_{rot}$ describing the relative populations of the two states, and calculated further the kinetic temperature $T_K$ of the gas following the updated conversion between $T_{rot}$ and $T_K$ given by Tafalla et al. (2004). We find $T_K = 20.1 \pm 0.8$\,K. Again, the FWHM resolution of these data is approximately 32\arcsec.

	We have also used the 450 $\mu$m and 850 $\mu$m data from the SCUBA Legacy Catalogues. The effective beam for the 450 $\mu$m data can be modeled as a combination of two Gaussian beams with FWHMs of 11\arcsec\ and 40\arcsec\ and relative intensities of 0.88 and 0.12, respectively, while the effective beam for the 850 $\mu$m data can be modeled as two Gaussian beams with FWHMs of 19\farcs5 and 40\arcsec\ and relative intensities of 0.88 and 0.12 respectively (Di Francesco et al. 2008). These effective beams are different from the intrinsic beam patterns of the JCMT due to post-processing performed during the compilation of the SCUBA Legacy Catalogues. We convolved the 850 $\mu$m image with the two-component model beam of the 450 $\mu$m data and we convolved the 450 $\mu$m image with the two-component model beam of the 850 $\mu$m data. Both data sets were then re-gridded onto our standard grid using a cubic spline interpolation routine. 
	
	Johnstone et al. (2006), for example, give the following relation for submillimeter dust continuum radiation: 
 \begin{equation} 
 \label{eqn:cont}
	 S(\nu) \propto \nu^\beta B_\nu(T_d) 
	 \end{equation}
where $S(\nu)$ is the observed flux, $\nu$ is the frequency, $B_\nu(T_d)$ is the Planck function evaluated at a temperature of $T_d$, $T_d$ is the dust temperature and $\beta$ is the dust emissivity spectral index. We have used Equation \ref{eqn:cont}, the ratio of 850 $\mu$m flux to 450 $\mu$m flux at the center of the N6 core and the NH$_3$ central temperature of 20.1 K to derive a value of 0.9 for $\beta$ at the center of the N6 core by assuming that the dust temperature is the same as the gas temperature. Models of the dense interiors of starless cores show that the dust and gas temperatures should be relatively similar (e.g. Hollenbach et al. 1991; Galli et al. 2002). A $\beta$ of 0.9 is comparable to values found in other star-forming regions (e.g. Beuther et al. 2004; Hogerheijde \& Sandell 2000; Goldsmith et al. 1997). It is, however, lower than the commonly used value of 2.0 in the ISM (Hildebrand 1983) and the value of 1.5 found for Oph A by Andr\'{e} et al. (1993). 

\begin{figure}[ht] 
   \centering
   \includegraphics[angle=-90, width=6in]{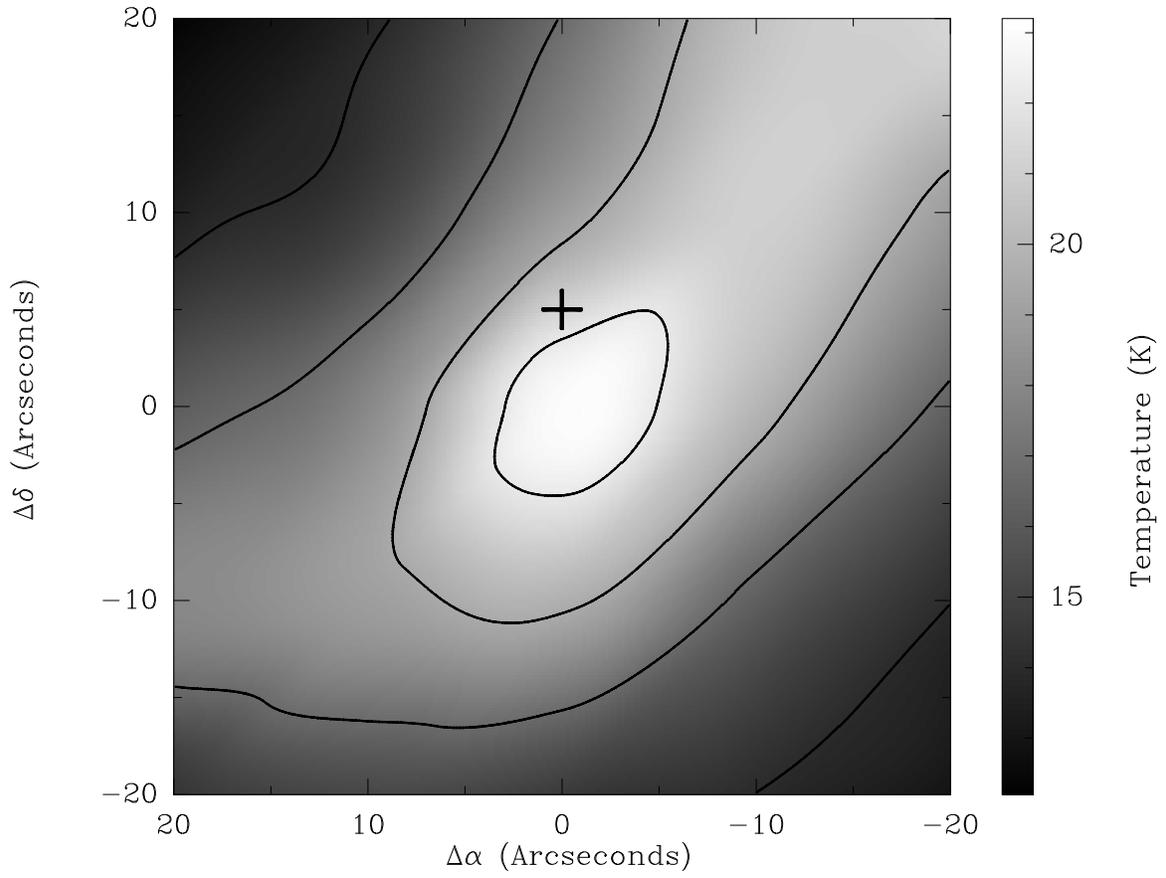} 
   \caption{Map of the dust temperatures derived from the submillimeter fluxes around the N6 core. The contours are 0.5 $\sigma$ contours where $\sigma$ = 4.5 K. The cross denotes the location of the submillimeter continuum peak found interferometrically by Bourke et al. (2009, in preparation).}
   \label{fig:temp}
\end{figure}

	Assuming that $\beta$ is constant over the N6 core, we were then able to derive dust temperatures for all points on our standard grid from the ratio of the submillimeter fluxes. Figure \ref{fig:temp} shows our derived dust temperature profile for the N6 core. With a $\beta$ of 0.9, we find that the central dust temperature is 20.2 K and that the average dust temperature is 15.4 K.  The dust temperature at the location of the N$_2$D$^+$ integrated intensity peak is 18.8 K.
	
	The SCUBA Legacy Catalogues also contain error maps for the submillimeter data. We recalculated two average dust temperatures for the N6 core after adding the 850 $\mu$m errors to the 850 $\mu$m data and subtracting the 450 $\mu$m errors from the 450 $\mu$m data and vice versa. By averaging the absolute differences between these new average dust temperatures and our original average dust temperature, we estimate that our continuum-derived dust temperatures are accurate to 4.5 K. Since the peak dust temperature of the N6 core is 4.8 K higher than the average dust temperature of the core, we believe that the dust temperature peak at the central location of the N6 core is significant, albeit marginally. It is also unlikely that noise would produce a coherent dust temperature peak that spans multiple pixels, as seen in Figure \ref{fig:temp}. A multi-wavelength analysis with high quality fluxes from at least four bands (and 350 microns in particular) should improve significantly the dust temperature characterization (see Schnee et al. 2007). We also consider our average dust temperature of 15.4 K to be in agreement with the average N$_2$H$^+$ excitation temperature derived by DAM04 (17.4 K), supporting their suggestion that the N$_2$H$^+$ 1\rto0 line is thermalized. 
			
	Di Francesco et al. (2008) suggested that the absolute fluxes in the 850 $\mu$m and 450 $\mu$m maps have uncertainties of 20\% and 50\% respectively. These errors are not expected to fluctuate significantly over small spatial scales, however. Therefore, these errors would only affect our determination of $\beta$ and/or the absolute temperature scale, but would not affect the derived temperature structure of the N6 core, since the relative temperature structure is independent of the choice of $\beta$. 

	The observed changes in the ratio of the submillimeter fluxes can be explained as due to either a temperature gradient, with a constant value of $\beta$ over the entire core, as assumed above, or due to a gradient in $\beta$, with a constant temperature everywhere in the core. If the entire N6 core were to be at a temperature of 20.1 K, $\beta$ would have to range from 0.9 at the center of the core to 0.23 at the periphery. Previous studies (e.g. Goldsmith et al. 1997) have found that $\beta$ decreases with increasing density and increasing grain size, the opposite of what would be needed in the N6 core for the core to be at a constant temperature. Thus, we believe that the submillimeter continuum data are indicative of a temperature gradient in N6 rather than a gradient in $\beta$.   	
								
\subsection{N$_2$D$^+$ Column Density and Deuteration Fraction}
\label{colden}
		
	Due to the blending of the N$_2$D$^+$ hyperfine components, we adopted the method given by Caselli et al. (2002) to determine the column densities of N$_2$D$^+$. This method assumes that the N$_2$D$^+$ emission is optically thin and in LTE. Since the N$_2$H$^+$ emission from the N6 core is only marginally optically thick (DAM04), we believe the assumption that the N$_2$D$^+$ emission is optically thin is reasonable. We used the dust temperatures calculated in Section \ref{temp}, to account for the decrease in temperature away from the N$_2$H$^+$ peak, and rejected any spectra that did not have a S/N greater than three. 

	The peak column density of N$_2$D$^+$ is $1.42 \times 10^{12}$ cm$^{-2}$ and occurs at the integrated intensity peak (at  [-10,10]). The ratio of N$_2$D$^+$ to N$_2$H$^+$ column density changes from approximately 0.01 at the N$_2$H$^+$ peak to approximately 0.02 at the N$_2$D$^+$ peak. 
 	
	We have assumed that the dominant uncertainties in calculating the N$_2$D$^+$ column densities come from the uncertainty in the temperatures, $\Delta T \approx 4.5$ K, and the uncertainty in the integrated intensities of the lines, 0.06 K km s$^{-1}$. We calculated new column densities using excitation temperatures 4.5 K greater than and less than our derived temperatures and integrated intensities 0.06 K km s$^{-1}$ greater than and less than our observed values. From this, we estimate the uncertainty in our column densities to be $\pm 2.7 \times 10^{11}$ cm$^{-2}$.

\subsection{Non-thermal Component of the Line Widths}
\label{nontherm}
	DAM04 found that the line widths for N$_2$H$^+$ in the N6 core were only slightly larger than the theoretical, thermal line widths. We have calculated the thermal line widths for the molecular transitions that we have observed using the gas temperature derived from the NH$_3$ observations, 20.1 K, for all of the transitions observed with the JCMT and the following equation from Myers et al. (1991):  
\begin{equation}
\Delta V_{thermal}=\sqrt{8ln(2)\times \frac{kT}{m}}
\end{equation}
where $\Delta V_{thermal}$ is the average thermal line width, $k$ is Boltzmann's constant, $T$ is the gas temperature, and $m$ is the molecular mass of the species. Since we are not concerned with pixel-to-pixel variations in line width for these calculations, we have chosen to use the 20.1K gas temperature derived from the NH$_3$ data. We also calculated the average non-thermal component of the line width for each transition by subtracting the thermal line width from the average observed line width in quadrature. For each transition, the average observed line width was calculated from the innermost 9 grid points. These results are summarized in Table \ref{tab:linewidth}. 

\section{DISCUSSION}
\label{discuss}
	
\subsection{Physical \& Chemical Structure of Oph N6} 
\label{structure}	

	At the high densities and low temperatures present at the center of starless cores, various molecules can freeze out onto dust grains and the abundances of these molecules, and their related ions, in the gas phase can become depleted (e.g., L\'{e}ger 1983; Bergin \& Tafalla 2007). It has also been suggested that the varying properties of molecules, such as their binding energies to dust grains and their reactivity with CO, will cause different molecules to deplete or form at different depths inside a core (Di Francesco et al. 2007 and references therein). This variable depletion creates shells of different chemical composition in the cloud (Tafalla et al. 2002; J\o rgensen et al. 2005).  

\begin{figure}[ht]
   \centering
   \includegraphics[angle=0, width=4in]{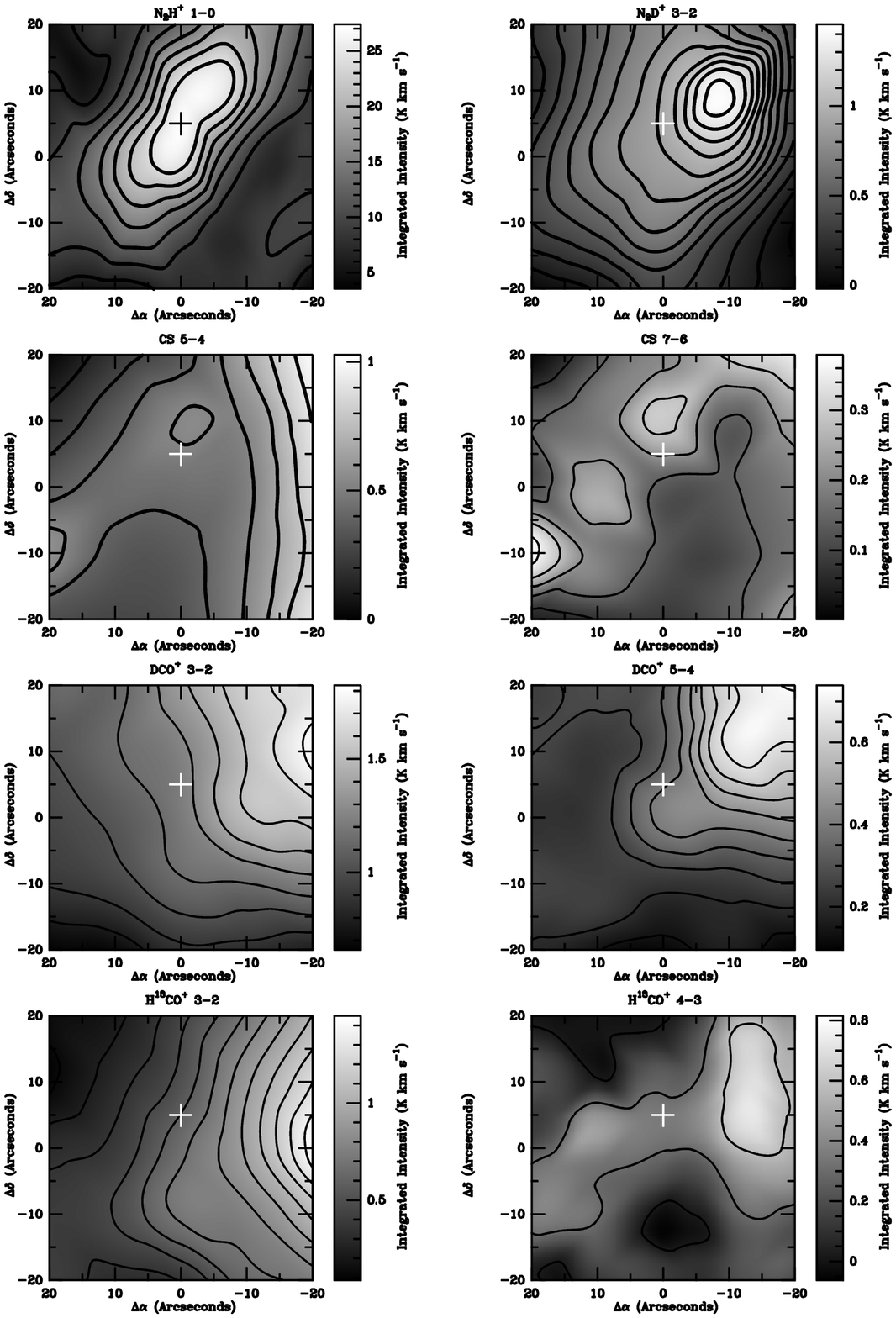} 
   \caption{From top left to bottom right, the integrated intensity maps of the N$_2$H$^+$ 1\rto0, N$_2$D$^+$ 3\rto2, CS 5\rto4, CS 7\rto6, DCO$^+$ 3\rto2, DCO$^+$ 5\rto4, H$^{13}$CO$^+$ 3\rto2 and H$^{13}$CO$^+$ 4\rto3 transitions. The cross on each map denotes the location of the submillimeter continuum peak found interferometrically by Bourke et al. (2009, in preparation). The contour lines are separated by 2 $\sigma$ on each map. The 1 $\sigma$ uncertainties for each map, in K km s$^{-1}$, from top left to bottom right, are 1.4, 0.03, 0.07, 0.04, 0.06, 0.03, 0.06 and 0.12 respectively. The uncertainty for the N$_2$H$^+$ was taken from DAM04, the uncertainties for the CS, DCO$^+$ and H$^{13}$CO$^+$ were found by multiplying the RMS from CLASS' baseline fit by the width of the lines and the uncertainty for the N$_2$D$^+$  was taken to be twice the product of the RMS of CLASS' baseline fit and the width of an N$_2$D$^+$ line.}
   \label{fig:allmol}
\end{figure}

	The localized emission peaks seen in the N$_2$H$^+$ and N$_2$D$^+$ integrated intensity maps, shown in Figure \ref{fig:allmol}, suggest that these molecules are present near the center of the core. Furthermore, the locations of the N$_2$H$^+$ and N$_2$D$^+$ emission peaks seem to suggest that the central part of the N6 core is a high-density ridge that stretches between [0,0] and [-10,10]. This is supported by the position of a small submillimeter continuum peak interferometrically detected by Bourke et al. (2009, in preparation) at [0,5]. 

	A comparison of Figure \ref{fig:temp} to Figure \ref{fig:allmol} shows that the strongest N$_2$D$^+$ emission occurs at a location with a colder dust temperature than the location of the peak N$_2$H$^+$ intensity. This is expected since colder temperatures tend to increase the chemical abundance of deuterated species in cores (Vastel et al. 2006). The colder temperature of the N$_2$D$^+$ maximum location would also explain why this was the location of DAM04's minimum observed, thermally dominated N$_2$H$^+$ line widths. While the D/H ratio of 0.02 observed is smaller than those seen in the most heavily deuterated cores (e.g., 0.1 to 0.5; Crapsi et al. 2005), it is consistent with the increased deuterium abundance seen in other starless cores (0.01 to 0.44; see Ceccarelli et al. 2007 and references therein) and is still a high level of fractionation.

	Figure \ref{fig:allmol} shows how the CS integrated intensity is smaller at the center of the N6 core than at its edges. This drop in CS integrated intensity suggests that the CS column density must also be significantly reduced at the center. As CS has been observed to deplete readily in other cold, dense cores (e.g., Bergin et al. 2001; Lai et al. 2003; Tafalla et al. 2006), we suggest that CS is freezing out onto dust grains in the center of N6. Thus, CS most likely only exists in the gas phase in an outer shell of material.

	Figure \ref{fig:allmol} shows also that DCO$^+$ and H$^{13}$CO$^+$ have a steady decreasing brightness gradient from the northwest edge to the center of our grid that is suggestive of a decline in the column densities of these molecules towards the center of N6.  Depletion of these molecules in cold, dense environments is also well known and has been observed in the L1498 and L1517B cores (e.g., Tafalla et al. 2006). The decrease in the integrated intensity of the CS lines at the center of the N6 core is more pronounced than in the DCO$^+$ or H$^{13}$CO$^+$ lines and thus, we suggest that CS is more heavily depleted towards the center and occurs in a shell-like structure further from the center than the H$^{13}$CO$^+$ and DCO$^+$. Since CO is known to destroy H$_2$D$^+$, the depletion of CO would be expected to lead to an increase in the rate of formation of deuterated species such as DCO$^+$ (Bergin \& Tafalla 2007 and references therein) and this may explain the differences in the DCO$^+$ and H$^{13}$CO$^+$ morphologies. Since HCO$^+$ is chemically similar to DCO$^+$ and H$^{13}$CO$^+$ we expect that the HCO$^+$ is depleting at a similar location to where the DCO$^+$ and H$^{13}$CO$^+$ are depleting. 
	
\begin{figure}[ht] 
   \centering
   \includegraphics[angle=0, width=8in]{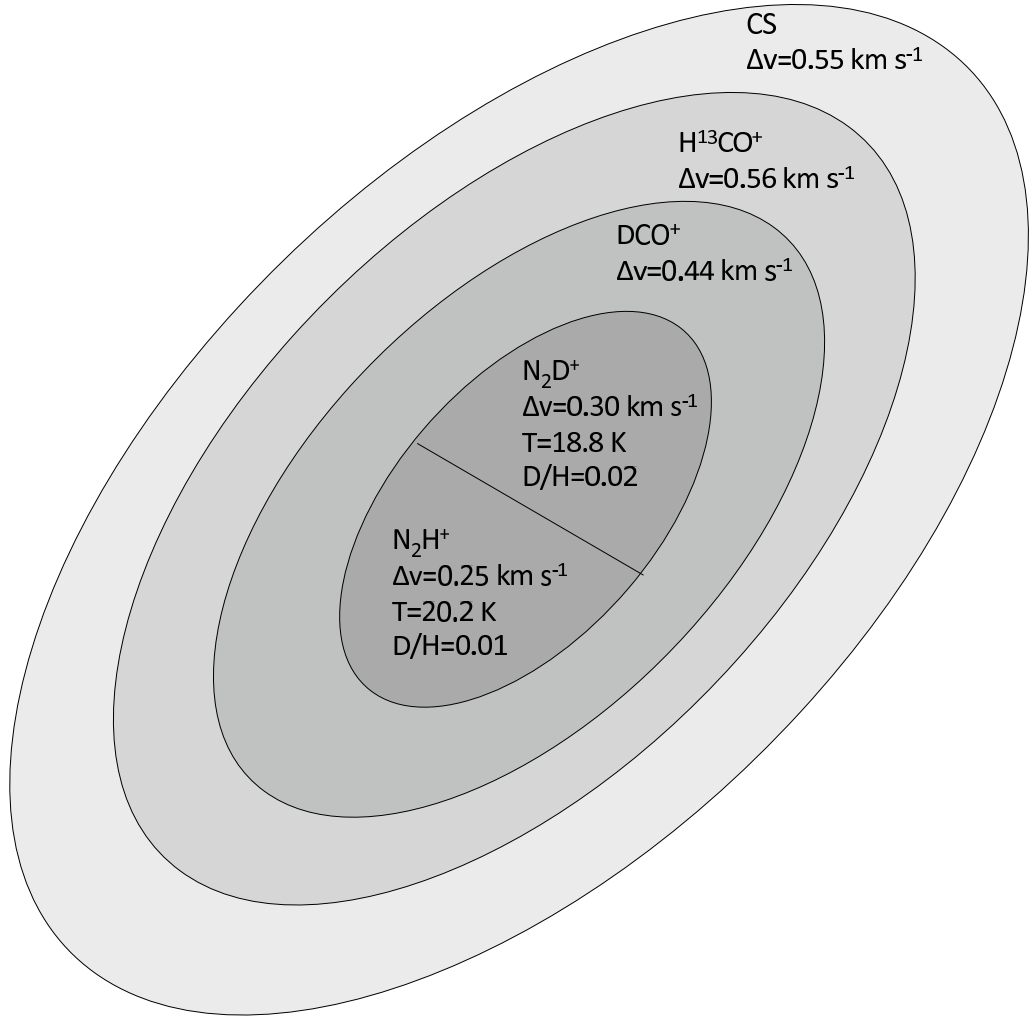} 
   \caption{Schematic diagram showing our model for the chemical structure of the N6 core.}
   \label{fig:structure}
\end{figure}

	Figure \ref{fig:structure} shows a cut-away diagram of the chemical shell structure that we suggest exists in the N6 core. For molecules where multiple transitions were observed, the combined average line width for these transitions is shown. 

\subsection{Turbulent Support}
\label{turbulence}

	Given the chemical structure of the N6 core described in Section \ref{structure}, the turbulent component of the line widths, (see Section \ref{nontherm} and Table \ref{tab:linewidth}), appears to increase with increasing distance from the center of the core. The only exception to this trend is that there is no significant difference in line width between the H$^{13}$CO$^+$ and CS.  The CS and H$^{13}$CO$^+$ line widths seem to be consistent with the line widths of the general Oph A medium (DAM04), which would explain why no further increase in line width is seen between CS and H$^{13}$CO$^+$. The slight increase in temperature towards the outer layers of the N6 core is insufficient to explain the significant increase in line width observed, as a temperature in excess of 200 K would be needed for the CS line widths to be entirely due to thermal broadening.

	The difference in line width between the N$_2$D$^+$ and the N$_2$H$^+$ may be due to the difference in beam sizes used in observing these molecular transitions. The larger 20\arcsec\ beam used to observe the N$_2$D$^+$ would have detected emission coming from more outlying, and thus more turbulent, gas than the approximately 10\arcsec\ x 6\arcsec\ beam used to observe the N$_2$H$^+$ would have. This more turbulent gas would have broadened the N$_2$D$^+$ lines slightly. This trend is supported by Andr\'{e} et al.'s (2007) finding of a 0.37 km s$^{-1}$ N$_2$H$^+$ average line width in the N6 core with a 26\arcsec\ beam. We also see a barely significant increase of 0.08 km s$^{-1}$ in average line width between the DCO$^+$ 5\rto4 transition, observed with a 14\arcsec\ beam, and the DCO$^+$ 3\rto2 transition, observed with a 20\arcsec\ beam. We see no significant difference between the line widths of the two observed transitions of CS and H$^{13}$CO$^+$, suggesting, again, that the CS and H$^{13}$CO$^+$ probe regions with the same level of turbulent motion as the general Oph A medium. 
 
	These results suggest that turbulent motions have been suppressed in the center of the N6 core and that the central part of the N6 core is no longer strongly interacting with the general, turbulent Oph A material. A possible mechanism for this dissipation of turbulent motion has been proposed by various authors (e.g., McKee et al. 1993; Nakano 1998; Myers 1998) who suggest that turbulent motions in dense cores could be reduced if the cores become smaller than the cut-off wavelength for MHD waves in the region. Myers (1998) suggested that these regions cut off from MHD waves could be on the order of 5000 AU ($\sim$ 40\arcsec \, at the distance of the N6 core) and DAM04 described how the N6 core seems to fit well the description of a kernel as described by Myers (1998). The suppression of turbulent motions and MHD waves in the N6 core also has a significant impact on the evolutionary state of the core, as it is believed that turbulence and magnetic fields are significant components of a core's support against gravitational collapse. The lack of such support in the N6 core could make the core prone to gravitational collapse.
	
\subsection{Infall in the N6 Core}
\label{infall}

	As first mentioned in Section \ref{morph}, many of the HCO$^+$ 3\rto2 spectra are double peaked and asymmetric, with the gap between the peaks occurring at the same frequency as the optically thin DCO$^+$ 3\rto2 line. If this dual peak structure were due to a second gas cloud in the line-of-sight, we would expect that the DCO$^+$ 3\rto2 spectra would appear at the same velocity as one of the two peaks in the HCO$^+$ spectra (e.g., Gregersen et al. 1997; Myers et al. 2000). Since the DCO$^+$ 3\rto2 spectral line appears between the two peaks in the HCO$^+$ spectra for all asymmetric HCO$^+$ spectra, this dual-peak structure is most likely due to self-absorption, rather than a second line-of-sight cloud. The rest frequencies of the HCO$^+$ 3\rto2 transition and DCO$^+$ 3\rto2 transition given by the JCMT and LAMDA databases are consistent and thus, we do not believe that this alignment is a coincidental artifact due to uncertainty in the rest frequencies of these lines. 
	 	
	Gregersen et al (1997) showed that the type of blue asymmetry predominately seen in the HCO$^+$ 3\rto2 spectra is indicative of infall motion. This asymmetry is most likely not due to random motions in the core, as this would produce a blue excess value close to zero. Rotation, alone, would also produce a blue excess value close to zero (Pavlyuchenkov et al. 2008). The prevalence of blue asymmetry suggests that the N6 core is collapsing. 

	Models of well-shielded, starless, dense cores have shown that the central kinetic temperature will be low ($\sim 10$ K) and that temperature increases with increasing radius due to heating by an external radiation field (e.g., Hollenbach et al. 1991).  After the onset of gravitational collapse, the central region should be warmed by the release of gravitational potential energy and the central temperature will increase (e.g., Lee et al. 2004).  This is the behavior seen in our dust continuum temperature map (Figure \ref{fig:temp}) in which the central region (20.2 K) is slightly warmer than the surrounding dust envelope (15.4 K) and is consistent with the onset of collapse of the core. As the temperature increase observed in the N6 core is still relatively small, it has probably only recently started to collapse. Gas at temperatures above 20.2 K, however, may exist in a region smaller than the size of our grid spacing. 
	
	Such a collapse is also consistent with the lack of turbulent support mentioned in Section \ref{turbulence}. The lack of CS emission from the center of N6 is also consistent with the core being at an early stage of collapse, since CS is predicted to evaporate off dust grains only during the later stages of gravitational collapse (e.g., Charnley 1997). The Spitzer c2d survey of the Oph A region did not detect any infrared source consistent with a young stellar object (YSO) within the boundaries of the N6 core, which further suggests that the N6 core has yet to evolve to the Class 0 protostar stage.

	It should be noted that the blue asymmetry in the HCO$^+$ spectra is not strongly peaked around the central parts of the N6 core but rather, is detected at multiple locations toward the periphery of the core as well as at the central locations of the core. Furthermore, since HCO$^+$ is most likely depleted in the innermost parts of the N6 core, as suggested by the depletion of H$^{13}$CO$^+$ and DCO$^+$, this infall signature is most likely only coming from the mid-to-outer layers of the cloud. Our tentative detection of a temperature increase at the center of the core, however, suggests that this infall motion does extend to the core's center. These points suggest that the N6 core is not undergoing an inside-out collapse, as predicted by the Shu (1977) model, but rather, is undergoing a global collapse along the lines of the Larson (1969) and Penston (1969) models. Indeed various numerical hydrodynamic simulations (e.g., Foster \& Chevalier 1993; Hennebelle et al. 2003) have shown that the Larson-Penston model provides the better approximation at early times of the collapse. We also cannot discount the possibility that we are just observing a global compression or contraction of the N6 core and not a gravitationally induced collapse. 
	
\section{CONCLUSIONS}
\label{conclusions}

	Our observations suggest that the Oph A-N6 core is an extremely young prestellar core, which has been recently cut off from MHD support and begun to collapse.  Evidence to support this claim comes from:
	
	\begin{enumerate}
\item {\it The chemical structure of N6} - The N$_2$D$^+$ map shows a well-defined core centered at [-10,10], which is slightly offset from the peak of the \diaz\ emission at [0,0] (DAM04). The N$_2$D$^+$ emission also appears to be slightly more compact than the \diaz\ emission, which shows a small ridge extending approximately 10$''$ to the northwest.  The DCO$^+$ and H$^{13}$CO$^+$ maps, however, show little sign of a centrally concentrated core-like structure. Rather, they mimic the dust continuum gradient (Di Francesco et al. 2008) suggesting that these two molecules are depleted at the center of the N6 core. The CS maps show a ring of emission around the central location of N6, also indicating significant depletion of CS at the center.  As a result, we believe that the N6 core is composed of shells of different chemical composition due to the different densities at which each chemical species freezes out onto dust grains.  In this model, the undepleted species N$_2$D$^+$ and \diaz\ trace the cold, high-density interior of N6, DCO$^+$ and H$^{13}$CO$^+$ trace a region which lies further from the center, and CS traces the outermost edges of the core. This type of molecular freeze out is a common feature in other dense starless cores (e.g. Tafalla et al. 2006). 

\item {\it The decreased line widths in the center of N6} - The average N$_2$D$^+$ and \diaz\ line widths in the central region of the core are $\approx$ 0.3 km s$^{-1}$.  The DCO$^+$ and CS line widths, however, are 0.44 km s$^{-1}$ and 0.55 km s$^{-1}$ respectively.  Given our proposed chemical structure for N6, this is consistent with the line widths increasing with increasing distance from the center.  A calculation of the non-thermal component required to match the observed line widths suggests that the non-thermal contribution to the line widths is greater at the edge of N6 than in the core center.  This supports the claim that the center of N6 has been cut off from turbulent support.

\item {\it Observed infall motion} - The HCO$^+$ spectra in N6 were found to exhibit distinct blue asymmetries across the N6 core.  The average asymmetry parameter for the HCO$^+$ spectra (Mardones et al. 1997) was found to be -0.71 which suggests that the mid-to-outer layers of the N6 core, which are probed by the optically thick HCO$^+$ spectra, are falling towards the center of the core. We also tentatively detected a slight increase in temperature towards the center of the N6 core, which suggests that this infall motion extends to the center of the core if this temperature increase is due to the release of gravitational potential energy. If the temperature is highest in the center and lower at the edges, the gradient of non-thermal motions is steeper than the line widths above suggest. 

\end{enumerate}

	While our data have revealed a shell-like structure for the N6 core, this analysis is still restricted by the relatively low spatial resolution of the JCMT at these frequencies. Interferometric observations of N6 of these transitions would help constrain the radius at which each species begins to deplete and may confirm our proposed chemical structure.  A detailed high-resolution examination of the kinematic structure of the N6 core is currently being conducted by Bourke et al. (2009, in preparation). An investigation into the ionization level of the N6 core (e.g. see Hezareh et al. 2009, in preparation; Plume et al. 1998) may also provide further evidence for whether MHD waves are not propagating into the interior of the core and hence, whether the cut-off from MHD waves is the cause of the decreased non-thermal support in the center of Oph A-N6.\\[0.5 in]

	We would like to thank the staff of the Joint Astronomy Centre in Hilo, David Gibson, Melissa Gurney, Tyler Bourke, Doug Johnstone, Helen Kirk, Sarah Sadavoy, Andrew Yau and Michael Weiser for their advice on this project. AP would also like to thank the Natural Sciences and Engineering Research Council of Canada for providing funding for this research via an Undergraduate Student Research Award and a Canada Graduate Scholarship. This research has made use of NASA's Astrophysics Data System.

\newpage

\newpage

\newpage

\newpage

\newpage

\newpage

\newpage

\newpage

\newpage

\begin{sidewaystable}[h]
\begin{small}
\begin{minipage}{\textwidth}
\caption{Observational parameters}
\begin{center}
\begin{tabular}{|p{1.3in}|p{0.8in}|p{0.8in}|p{0.8in}|p{0.8 in}|p{0.7 in}|p{0.8 in}|p{0.8 in}|p{0.9in}|}
\hline
Transition & Frequency (GHz)$^a$ & Front End Instrument & Back End Instrument & Spectral Resolution (km s$^{-1}$) & FWHM of beam (\arcsec) & Beam Efficiency & Baseline RMS (K)& Critical Density ($10^6$ cm$^{-3}$)$^b$\\
\hline
N$_2$D$^+$ J = 3\rto2 & 231.321965 & RxA3 & DAS$^c$ & 0.1 & 20 & 0.77 & 0.08 & 3.0$^d$ \\
\hline
DCO$^+$ J = 3\rto2 & 216.112604 & RxA3 & DAS & 0.1 & 20 & 0.77 & 0.12 & 1.8  \\
\hline
DCO$^+$ J = 5\rto4 & 360.169881 & RxB3 & DAS & 0.13 & 14 & 0.70 & 0.08 & 9.2 \\
\hline
H$^{13}$CO$^+$ J = 3\rto2 & 260.255478 & RxA3 & ACSIS$^e$ & 0.035 & 20 & 0.69 & 0.10 & 3.2 \\
\hline
H$^{13}$CO$^+$ J = 4\rto3 & 346.99854 & HARP$^f$ & ACSIS & 0.025 & 14 & 0.63 & 0.28 & 8.2\\
\hline
CS J = 5\rto4&  244.935643 & RxA3 & ACSIS & 0.035 & 20 & 0.69 & 0.10 & 8.1\\ 
\hline
CS J = 7\rto6 & 342.883000 & RxB3 & DAS & 0.13 & 14 & 0.70 & 0.06 & 25 \\ 
\hline
HCO$^+$ J = 3\rto2 & 267.557619 & RxA3 & DAS & 0.1& 20 & 0.77 & 0.10 & 3.5  \\
\hline
NH$_3$ (1,1) & 23.694495 & GBT K-band (upper) & GBT spectrometer & 0.077 & 32 & 0.78 & 0.02 & 0.0021 \\  
\hline
NH$_3$ (2,2) & 23.722633 & GBT K-band (upper) & GBT spectrometer & 0.077 & 32 & 0.78 & 0.02 & 0.0021 \\
\hline
\end{tabular}
\end{center}
\label{tab:obs}
\footnotetext[1]{All frequencies were taken from the JCMT database.}
\footnotetext[2]{Calculated from the data in the Leiden Atomic and Molecular Database (Sch\"{o}ier et al. 2005).}
\footnotetext[3]{Digital Autocorrelation Spectrometer.}
\footnotetext[4]{Assumed to be equal to the critical density of the N$_2$H$^+$ J = 3\rto2 transition.} 
\footnotetext[5]{Auto-Correlation Spectrometer and Imaging System.}
\footnotetext[6]{Heterodyne Array Receiver Programme.}
\end{minipage}
\end{small}
\end{sidewaystable}

\newpage

\begin{table}[h]
\begin{minipage}{\textwidth}
\caption{Line properties for the inner nine pointings}
\begin{center}
\begin{tabular}{|p{1.5in}|p{0.8in}|p{0.8in}|p{0.8in}|p{0.9 in}|}
\hline
Transition & $<\Delta V>$ (km s$^{-1}$) & $\delta$ $<\Delta V>$ (km s$^{-1}$)& $<V_{LSR}>$ (km s$^{-1}$) & $\delta$ $<V_{LSR}>$ (km s$^{-1}$)\\
\hline
N$_2$D$^+$ J = 3\rto2 & 0.30 & 0.04 & 3.49 & 0.02 \\
\hline
DCO$^+$ J = 3\rto2 & 0.48 & 0.06 & 3.43 & 0.06 \\
\hline
DCO$^+$ J = 5\rto4 & 0.40 & 0.05 & 3.51 & 0.05 \\
\hline
H$^{13}$CO$^+$ J = 3\rto2 & 0.58 & 0.07 & 3.62 & 0.08 \\
\hline
H$^{13}$CO$^+$ J = 4\rto3 & 0.54 & 0.12 & 3.58 & 0.08 \\
\hline
CS J = 5\rto4& 0.55 & 0.10 & 3.58 & 0.12 \\ 
\hline
CS J = 7\rto6 & 0.54 & 0.21 & 3.62 & 0.07 \\ 
\hline
\end{tabular}
\label{tab:lineprop}
\end{center}
\end{minipage}
\end{table}

\newpage
 
\begin{table}[h]
\begin{minipage}{\textwidth}
\caption{Average line widths for the nine most central spectra in our standard grid}
\begin{center}
\begin{tabular}{|p{1.5in}|p{0.7in}|p{0.9in}|p{1.0in}|}
\hline
Transition & $\Delta V_{obs}$ (km s$^{-1}$) & $\Delta V_{thermal}$ (km s$^{-1}$) & $\Delta V_{non-thermal}$ (km s$^{-1}$)\\
\hline
N$_2$H$^+$ J= 1\rto0 & 0.25 & 0.18 & 0.17\\
\hline
N$_2$D$^+$ J = 3\rto2 & 0.30 & 0.18 & 0.24 \\
\hline
DCO$^+$ J = 3\rto2 & 0.48 & 0.18 & 0.45 \\
\hline
DCO$^+$ J = 5\rto4 & 0.40 & 0.18 & 0.36 \\
\hline
H$^{13}$CO$^+$ J = 3\rto2 & 0.58 & 0.18 & 0.56 \\
\hline
H$^{13}$CO$^+$ J = 4\rto3 & 0.54 & 0.18 & 0.51 \\
\hline
CS J = 5\rto4& 0.55 & 0.15 & 0.53 \\ 
\hline
CS J = 7\rto6 & 0.54 & 0.15 & 0.52 \\ 
\hline
\end{tabular}
\label{tab:linewidth}
\end{center}
\end{minipage}
\end{table}

\end{document}